\documentclass[letter]{jpsj2}

\def\beq{\begin{equation}}
\def\eeq{\end{equation}}
\newcommand{\mbold}[1]{\mbox{\boldmath $ #1 $}}

\usepackage{color}

\title{%
Nonmonotonic Relaxation in Systems with Reentrant-Type Interaction 
}

\author{%
Seiji Miyashita\thanks{E-mail address: miya@spin.phys.s.u-tokyo.ac.jp }$^{1,2}$,
Shu Tanaka\thanks{E-mail address: shu-t@spin.phys.s.u-tokyo.ac.jp}$^{1,2}$, 
and  
Masaki Hirano\thanks{E-mail address: hirano@spin.phys.s.u-tokyo.ac.jp}$^{1,2}$ 
}

\inst{%
$^1$ Department of Physics, The University of Tokyo
7-3-1 Hongo, Bunkyo-ku, Tokyo 113-0033 Japan. \\
$^2$ CREST, JST, 4-1-8 Honcho Kawaguchi, Saitama 332-0012, Japan.
}

\recdate{\today}

\abst{%
Recently, interesting nonmonotonic time evolution has been pointed out 
in the experiments by J\"onsson {\it et al.} and Jonsson {\it et al.}
and also in the numerical simulation by Takayama and Hukushima, where
the magnetic susceptibility does not monotonically relax to the equilibrium value, 
but moves to the opposite side. 
We study the mechanism of this puzzling nonmonotonic dynamical property in a frustrated Ising model
in which the equilibrium correlation exhibits nonmonotonic temperature dependence (reentrant type). 
We study the time evolution of the spin correlation function after a sudden change of temperature.  
We find that the value of the correlation function shows nonmonotonic relaxation,
and we analyze the mechanisms of the nonmonotonicity.
We also point out that competition between different configurations generally causes nonmonotonic
relaxation.
}

\kword{%
frustration, decoration bond, reentrant phase transition, relaxation, Ising spin system 
}

\begin{document}
\maketitle


Recently, Takayama and Hukushima pointed out an interesting dynamical property of the magnetization
of Ising spin glasses in the magnetic field after the halt of the field cooling\cite{takayama}. 
They studied the dynamics of magnetization at various temperatures and in various field protocols.
That is, they decreased the temperature at several speeds in a finite magnetic field, and
observed the change in magnetization over time. 
In particular, they found that the magnetization after the halt of field cooling  
beyond the spin glass transition temperature sometimes shows nonmonotonic relaxation. 
In some cases, the magnetization at the end of the field cooling process was smaller than the
equilibrium value. 
However, the magnetization still decreased
which was the opposite tendency to the equilibrium value.
The observation is consistent with the experiments
of the field-cooled protocol in Fe$_{0.5}$Mn$_{0.5}$TiO$_3$\cite{Tjonsson} and
Fe$_{0.55}$Mn$_{0.45}$TiO$_3$,\cite{PEjonsson} 
which are good model systems for a short-range interacting Ising spin glass.
This type of nonmonotonic dynamics in the relaxation toward the equilibrium state is interesting,
and in this Letter we study possible mechanisms using simplified models.

In frustrated spin systems, the correlation often shows nonmonotonic dependence
on the temperature.\cite{syozi,miya1} 
This nonmonotonic dependence causes reentrant phase transitions\cite{miya1,kitatani,Jascur}, and also
can be an origin of temperature chaos in spin glasses.\cite{vincent,tanaka} 
In a simple ferromagnetic model, a phase transition occurs between the high-temperature paramagnetic phase 
and the ferromagnetic phase.
The ferromagnetic state is energetically favorable and realized at low temperature. 
This situation gives a standard order-disorder phase transition. 
In contrast, in frustrated systems, different 
types of ordered states are often nearly degenerate. 
In such cases, different order phases are realized at different temperatures. 
For example, when the temperature decreases from a high temperature, 
the system may show successive phase transitions, i.e., 
from the paramagnetic to an ordered phase 
(e.g., with an antiferromagnetic order) and then to another ordered phase 
(e.g., with a ferromagnetic order). 
Generally, this type of successive phase transition is called a reentrant
phase transition, although originally ``reentrant'' means the phase transitions from
paramagnetic phase to an ordered phase and then to the paramagnetic phase again. 
The nature of reentrant phase transitions has been studied exactly using 
the two-dimensional Ising model.\cite{kitatani}
This nonmonotonic dependence is not only important in the phase transition 
but also plays an important role in the temperature dependence of the 
short-range correlations.
In frustrated systems such as spin glasses, various 
frustrated local interaction configurations are realized. 
In such systems, we expect 
the local correlations to show nonmonotonic temperature dependence, and we expect the
ordered structure to change globally with temperature, which can be a possible mechanism of
the temperature chaos and the rejuvenation phenomena.\cite{vincent}

\begin{figure}[h]
$$\begin{array}{ccc}
\includegraphics[height=20mm]{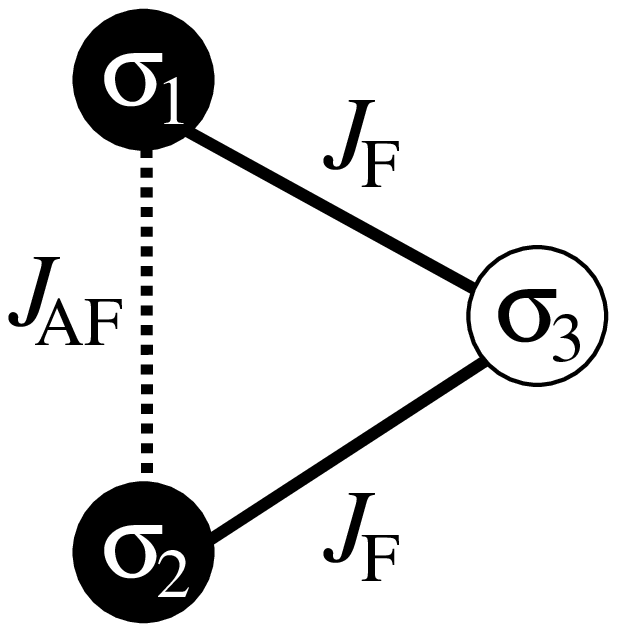}&\hspace{5mm}&
\includegraphics[height=20mm]{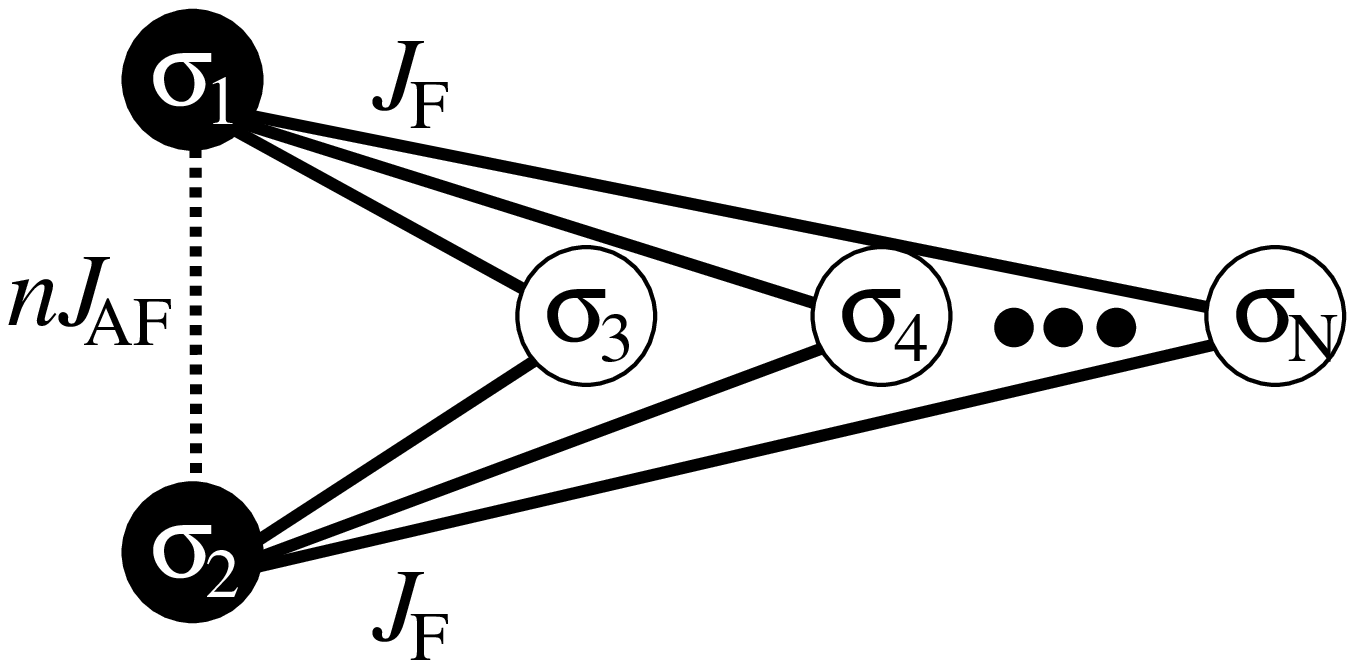}\\
({\rm a}) && ({\rm b}) \end{array}
$$
\caption{(a) A frustrated lattice. The dotted line denotes $J_{\rm AF}$ and the solid lines denote
$J_{\rm F}$. The effective coupling between $\sigma_1$ and $\sigma_2$ is given by
$K^{(1)}_{\rm eff}(T)$.
(b) A frustrated lattice in which the structure of Fig.~1(a) is used $n$ times.
The effective coupling in this structure is $K^{(n)}_{\rm eff}(T)$.}
\label{fig-lattice}
\end{figure}

In this Letter, we study the dynamics of the correlation function in a frustrated Ising model
\beq
{\cal H}=\sum_{ij}J_{ij}\sigma_i\sigma_j-H\sum_i\sigma_i,
\eeq
which shows reentrant-type temperature dependence of the correlation function.
Because of the freedom of the local gauge transformation 
$\{\sigma_i\rightarrow -\sigma_i, J_{ij}\rightarrow -J_{ij}\}$,
we can have many configurations of interactions with the same property of frustration. 
We call this degeneracy ``Mattis degeneracy''. 
In the case of $H=0$, all the models in the Mattis degeneracy have the same thermodynamic
properties.
On the other hand, the magnetic response, of course, depends on the configuration $\{J_{ij}\}$.
Thus, we mainly study the correlation function of spins, which is essentially the same (apart from
the sign) in the Mattis degenerate systems.

\begin{figure}[h]
\begin{center}
\includegraphics[height=45mm]{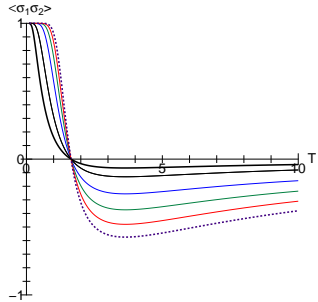}
\end{center}
\caption{(Color online) Temperature dependence of the correlation function
$\left\langle \sigma_1 \sigma_2 \right\rangle$ 
for $n=$1, 2, 4, 6, 8, and 10.
The bold solid curve denotes the data for $n=1$, and the bold dotted line denotes the data for $n=10$.
The intermediate lines show $n$=2, 4, 6, and 8.}
\label{fig-Keff}
\end{figure}

One of the simplest models of this type is depicted in Fig.~\ref{fig-lattice}(a).
The Hamiltonian is given by
\beq
{\cal H}=J_{\rm AF}\sigma_1\sigma_2 -J_{\rm F}\sigma_3(\sigma_1+\sigma_2),
\label{ham1}
\eeq  
where $2J_{\rm F}>J_{\rm AF}>0$. The effective coupling $K_{\rm eff}(T)$ 
between the spins $\sigma_1$ and $\sigma_2$ at temperature $T$
is defined by
\beq
\sum_{\sigma_3=\pm 1}e^{-\beta{\cal H}}=A(T)e^{K^{(1)}_{\rm eff}(T)\sigma_1\sigma_2}
\eeq
and effective coupling between $\sigma_1$ and $\sigma_2$ is 
\beq
K^{(1)}_{\rm eff}(T)=-{J_{\rm AF}\over k_{\rm B}T}
+{1\over2}\log\left(\cosh\left({2J_{\rm F}\over k_{\rm B}T}\right)\right),
\eeq
where $A(T)$ is an analytic function of $T$, and the equilibrium correlation 
$\left\langle \sigma_1 \sigma_2 \right\rangle$ is given by $\tanh K_{\mathrm{eff}}$.
In Fig.~\ref{fig-Keff}, we plot the equilibrium correlation 
$\left\langle \sigma_1 \sigma_2 \right\rangle$ 
as a function of the temperature for the parameters $J_{\rm F}=1$ and $J_{\rm AF}=0.5$.
The bold solid curve denotes the case of the model given by eq.~(\ref{ham1}).
Hereafter, we take $J_{\rm F}$
as the unit of energy. The temperature is also scaled by $J_{\rm F}$.
In Fig.~\ref{fig-Keff}, we can observe nonmonotonic temperature dependence. The correlation 
$\left\langle \sigma_1 \sigma_2 \right\rangle$ at the higher-temperature side is
small. To increase this amplitude, we provide a multiplied decoration bond
with $n$ intermediate spins (the open circles depicted in
Fig.~\ref{fig-lattice}(b)). 
Therefore, the total number of spins is $N=n+2$. 
The Hamiltonian in this case is given by
\beq
{\cal H}^{(n)}=nJ_{\rm AF}\sigma_1\sigma_2 -J_{\rm F}(\sum_{k=3}^{n+2}\sigma_k)(\sigma_1+\sigma_2),
\eeq  
and the
 effective coupling between $\sigma_1$ and $\sigma_2$ is given by
\beq
K^{(n)}_{\rm eff}(T)=n K^{(1)}_{\rm eff}(T)=
-n{J_{\rm AF}\over k_{\rm B}T}
+{n\over2}\log\left(\cosh\left({2J_{\rm F}\over k_{\rm B}T}\right)\right).
\eeq
The correlation functions for various values of $n$ are also plotted in Fig.~\ref{fig-Keff}.

Now we study the dynamics of the correlation function 
\beq
C(t)=\sum_{\{\sigma_i=\pm 1\}}P(\{\sigma_i\},t)\sigma_1\sigma_2,
\eeq
where $P(\{\sigma_i\},t)$ is the distribution function at time $t$. We adopt the Glauber-type
kinetic Ising model\cite{glauber} for the time evolution
$${\partial P(\sigma_1,\cdots,\sigma_i,\cdots,\sigma_N,t)\over \partial t}$$
\beq
\label{eq:master}
=-\sum_iP(\sigma_1,\cdots,\sigma_i,\cdots,\sigma_N,t)w_{\sigma_i\rightarrow -\sigma_i}
+\sum_iP(\sigma_1,\cdots,-\sigma_i,\cdots,\sigma_N,t)w_{-\sigma_i\rightarrow \sigma_i},
\eeq
with the transition probability per unit time
\beq
w_{\sigma_i\rightarrow -\sigma_i}={P_{\rm eq}(\sigma_1,\cdots,-\sigma_i,\cdots,\sigma_N)\over
 P_{\rm eq}(\sigma_1,\cdots, \sigma_i,\cdots,\sigma_N)
+P_{\rm eq}(\sigma_1,\cdots,-\sigma_i,\cdots,\sigma_N)},
\eeq
where 
\beq
P_{\rm eq}(\sigma_1,\cdots,\sigma_i,\cdots,\sigma_N)=
{e^{-\beta{\cal H}(\{\sigma_i\})}\over Z},\quad Z={\rm Tr} e^{-\beta{\cal H}(\{\sigma_i\})}.
\eeq
It is convenient to use the vector $\mbold{P}(t)$ consisting of the probabilities of the states
\beq
\mbold{P}(t)=\left(\begin{array}{c} P(++,\cdots,+++,t) \\
  P(++,\cdots,++-,t)\\ \vdots \\  P(--,\cdots,---,t)\end{array} \right).
\eeq
The dynamics is expressed by
\beq
\mbold{P}(t+\Delta t)={\cal L}\mbold{P}(t),
\label{eq:Leq}
\eeq
where ${\cal L}$ is a $2^N\times 2^N$ matrix with matrix elements 
at the small limit of $\Delta t$.
\beq
{\cal L}_{ij}={1\over N}w_{j\rightarrow i}\Delta t \quad {\rm for} \quad i\ne j
\eeq
and 
\beq
{\cal L}_{ii}=1-\sum_{j\ne i}{\cal L}_{ji}. 
\eeq

\begin{figure}[h]
\begin{center}
\includegraphics[height=45mm]{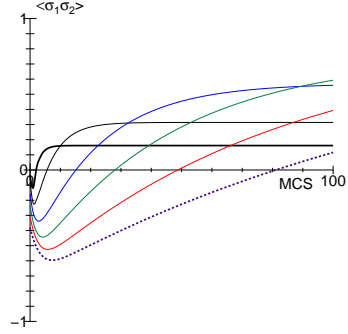}
\end{center}
\caption{(Color online) Time (Monte Carlo step) dependence of the correlation function for $n$ = 2, 4, 6, 8, and 10.
The types of line mean the same as those in Fig.~\ref{fig-Keff}.}
\label{fig-time}
\end{figure}

Hereafter, we adopt eq.~(\ref{eq:Leq}) 
as an approximate discrete form of eq.~(\ref{eq:master}).
Here, we suddenly change the temperature from $T_1=10$ to $T_2=1$
and study the subsequent dynamics by iterating the time evolution operator ${\cal L}$ with
$\Delta t=1$. 
We call an update by this procedure a ``Monte Carlo step (MCS)''. 
The initial probability distribution is set to be the equilibrium one at $T=T_1$. 

In Fig.~\ref{fig-time}, we depict the time evolution of the correlation $C(t)$.
We find that $C(t)$ first decreases, 
the amplitude increases on the negative side,
it then becomes positive, and finally it reaches the equilibrium value at $T=T_2$.
This observation indicates that even if the temperature is changed suddenly to $T_2$,
the correlation function does not necessarily relax directly toward its new equilibrium value,
but it can show nonmonotonic relaxation.
We may understand that the ``effective temperature'' of the system decreases gradually even when the
temperature of the thermal bath is changed suddenly. The correlation function, as well as other
quantities, shows a similar dependence to its temperature dependence. 
That is, if a quantity shows a nonmonotonic 
temperature dependence in equilibrium, it tends to show nonmonotonic relaxation.

To understand this nonmonotonic behavior, 
we analyze the time evolution from the viewpoint of the
eigenvalue problem of the time-evolution operator ${\cal L}$.\cite{eigen}
Let $\mbold{u}_k$ and $\lambda_k$ be the $k$th right eigenvector 
of ${\cal L}$ and the eigenvalue, respectively ($k=1,\cdots,2^N$)
\beq
{\cal L}\mbold{u}_k=\lambda_k\mbold{u}_k.
\eeq
Here, we assume that $1=\lambda_1> \lambda_2\ge \lambda_3,\cdots,\ge \lambda_{2^N}$.
The state $\mbold{u}_1$ gives the equilibrium state at $T=T_2$.
The initial state (the equilibrium state at $T=T_1$) is expanded using $\{\mbold{u}_k\}$ as
\beq
\mbold{P}(0)=\mbold{u}_1+\sum_{k=2}^{2^N}c_k\mbold{u}_k.
\eeq 
After $t$ time evolutions by ${\cal L}$, the state evolves to
\beq
\mbold{P}(t)={\cal L}^t\mbold{P}(0) =\mbold{u}_1+\sum_{k=2}^{2^N}c_k\lambda_k^t\mbold{u}_k.
\eeq 
The contribution of the $k$th mode to the correlation function is
\beq
C_k(t)=c_k\lambda_k^t\sum_{\{\sigma_i=\pm 1\}}{u}_k(\{\sigma_i\}) \sigma_1\sigma_2.
\eeq
The sign of $c_k$ can be either positive or negative, and the contribution of each
mode $C_k(t)$ causes a decrease or increase in $C(t)$ corresponding to the sign
of $c_k\sum_{\{\sigma_i=\pm 1\}}{u}_k(\{\sigma_i\}) \sigma_1\sigma_2$.
The relaxation time of each mode is different and $C(t)$ can be nonmonotonic.
Here we demonstrate this situation taking the simplest case of $n=1$.

\begin{figure}[h]
$$\begin{array}{cc}
\includegraphics[height=35mm]{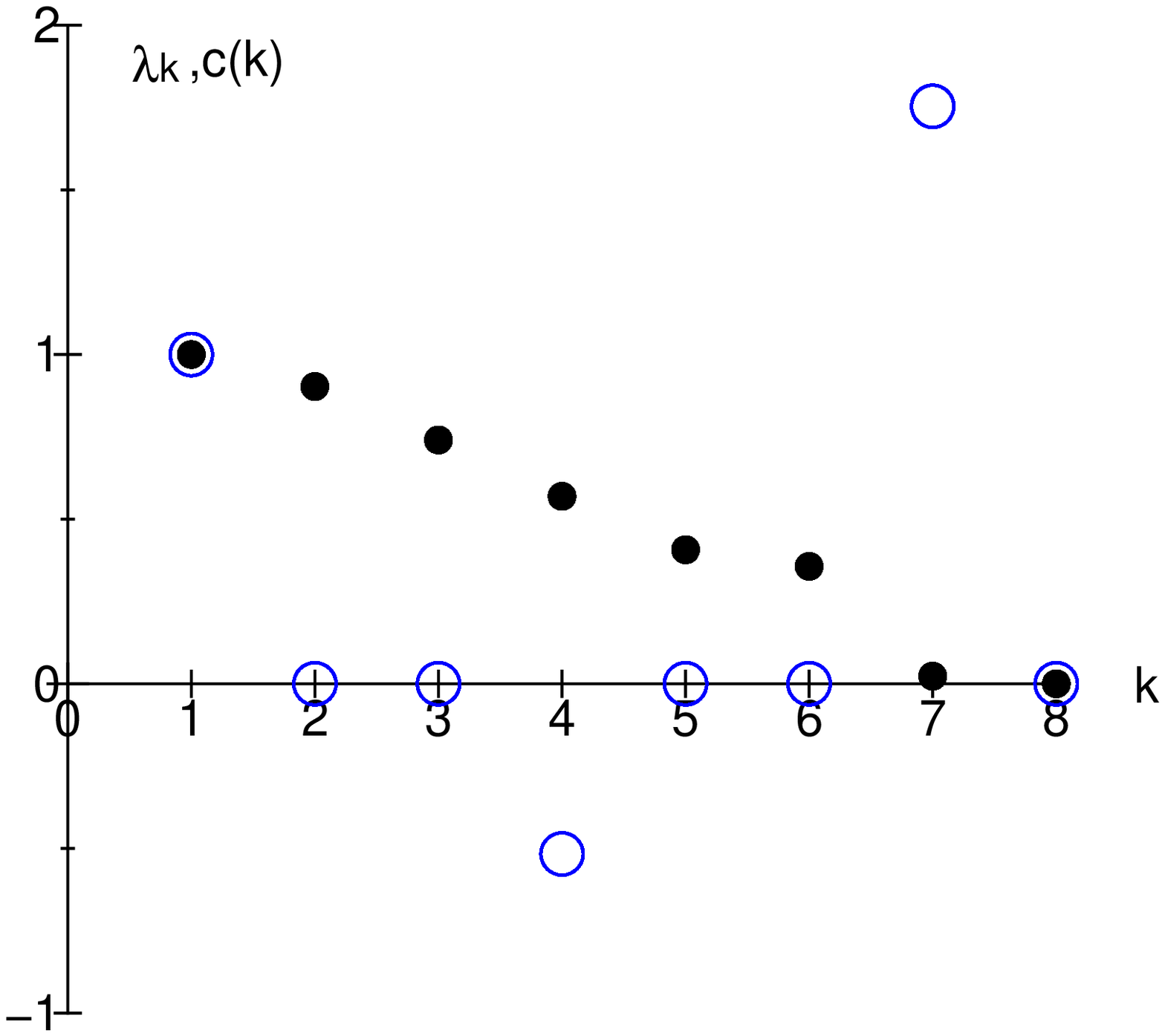}&
\includegraphics[height=35mm]{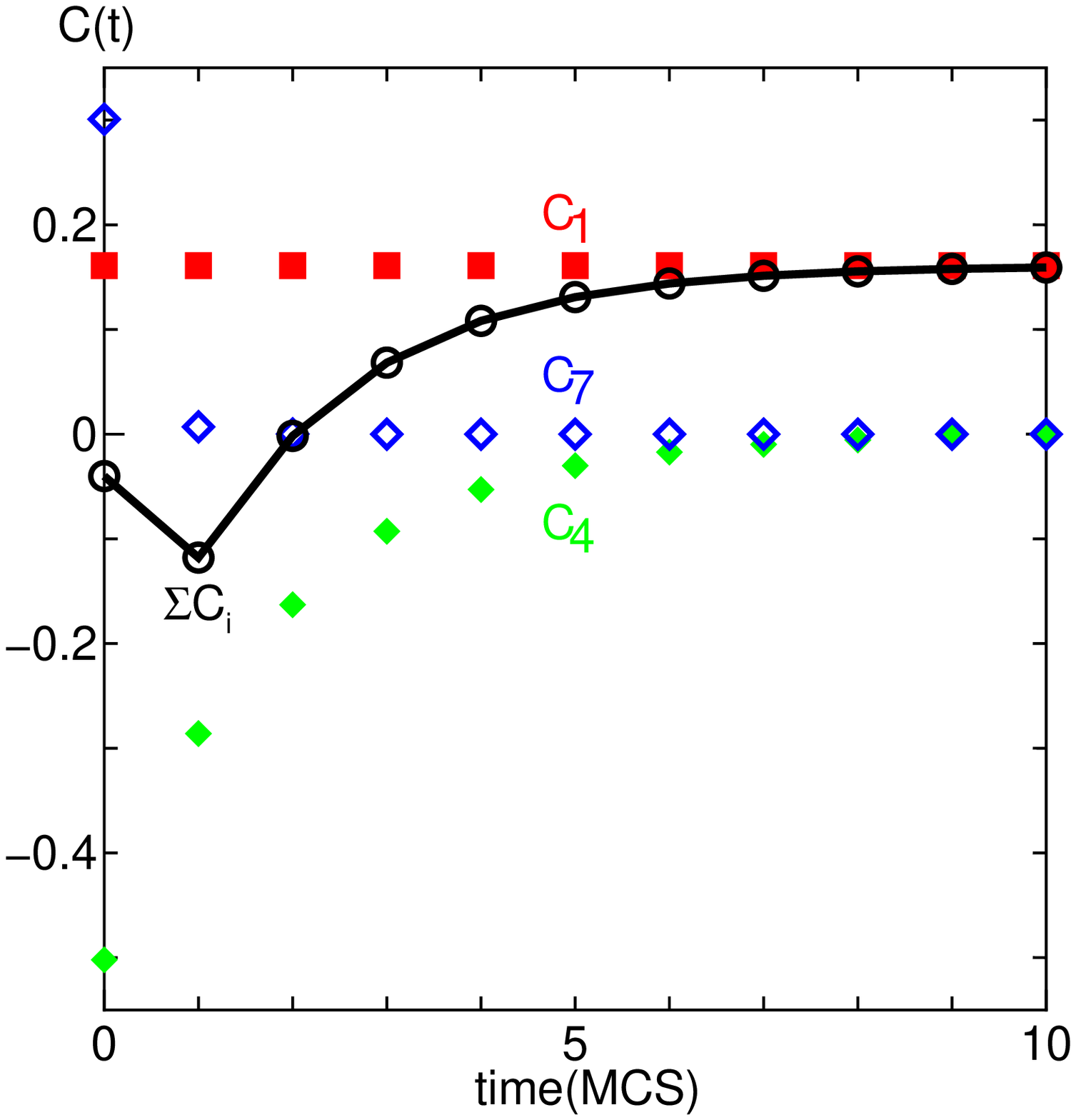}\\
({\rm a}) & ({\rm b})\end{array}$$
\caption{(Color online)
(a) Eigenvalues $\lambda_k$ for the case of $n=1$ (solid circles) of the time evolution operator at $T=T_2(=1)$, 
and 
the coefficients $c_k$ (open circles) for the equilibrium state at $T=T_1(=10)$.
Only the modes of $k=$ 4 and 7 give contributions.
(b) Time evolution of $C(t)$ (bold curve) with the contributions from the
4th mode (solid diamonds) and 7th mode (open diamonds).
Open circles correspond to the sum of $c_i$.}
\label{fig-eigen}
\end{figure}

In the case of $n=1$, there are 8 modes.  
The eigenvalues $\lambda_k$ and the coefficients $c_k$
are plotted in Fig.~\ref{fig-eigen}(a). 
We take the initial state to be the equilibrium state at $T_1$, 
and the contribution from some modes is zero because of the symmetry.
For example, modes that are antisymmetric in exchanging sites 1 and 2 do not contribute. 
In the present case, only two modes $k=4$ and 7 contribute.
The changes in $C_4(t)$ and $C_7(t)$ are depicted in Fig.~\ref{fig-eigen}(b).
The contribution from the 7th mode with a short relaxation time (open diamonds) 
is positive and it relaxes rapidly. This causes the decrease in $C(t)$ in the early stages.
On the other hand, the contribution of the slow-relaxing mode ($k=4$) is negative and it
causes an increase in $C(t)$ to the equilibrium value. 
For larger $n$, we found similar behavior (not shown).

\begin{figure}[h]
\begin{center}
\includegraphics[height=45mm]{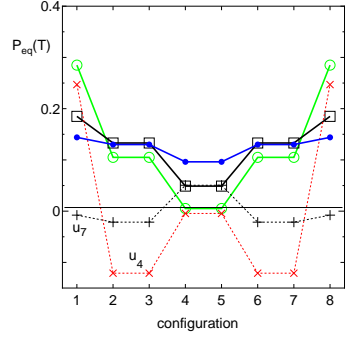}
\end{center}
\caption{(Color online)
Probability distributions of the equilibrium states at $T=1$ (open circles), 
2 (squares), and 10 (closed circles), 
and eigenrelaxation modes $\mbold{u}_4$ (crosses) and $\mbold{u}_7$ (pluses).
Lines between points are drawn to highlight the trends. 
The configurations $1, 2, 3, 4,\cdots,8$ denote the states 
$(\sigma_3,\sigma_2,\sigma_1)=(+++),(++-),(+-+),(+--)\cdots,(---)$, respectively.}
\label{fig-mode}
\end{figure}

It is interesting to note the characteristics of the relaxation modes.
In Fig.~\ref{fig-mode}, we plot the probability distribution in the phase space of
the equilibrium states $\mathbf{P}_{\rm eq}(T)$ at $T=1$, 2, and 10.
The configurations $1, 2, 3, 4,\cdots, 8$ denote $\left( \sigma_3, \sigma_2, \sigma_1 \right)
= \left( + + + \right)$, $\left(+ + - \right)$, $\left(+ - + \right)$, 
$\left(+ - - \right)$, $\cdots$ $\left(- - - \right)$, respectively. 
In configurations 1, 4, 5, and 8, 
the spins $\sigma_1$ and $\sigma_2$ have the same sign and 
thus there is a ferromagnetic correlation between $\sigma_1$ and $\sigma_2$,
and the configurations 2, 3, 6, and 7 have an antiferromagnetic correlation.
We also plot the eigenvectors $\mbold{u}_4$ and $\mbold{u}_7$ at $T=T_2(=1)$. 
There, we find that the fast relaxation mode $\mbold{u}_7$ 
has large amplitudes at configurations 4 and 5,
and the reduction of this mode causes the change from
$\mathbf{P}_{\rm eq}(10)$ to $\mathbf{P}_{\rm eq}(2)$, where the probabilities of 
the energetically unfavorable configurations 4 and 5 are reduced, 
and the antiferromagnetic correlation between $\sigma_1$ and $\sigma_2$ slightly increases. 
The slow relaxation mode $\mbold{u}_4$ corresponds to the difference between
$\mathbf{P}_{\rm eq}(2)$ and $\mathbf{P}_{\rm eq}(1)$ 
where the ferromagnetic correlation between $\sigma_1$ and $\sigma_2$ 
increases.
Here we find that the relaxation modes reflect the temperature dependence of the equilibrium state.
This fact corresponds to the gradual cooling of the system.

\begin{figure}[h]
$$\begin{array}{cc}
\includegraphics[height=35mm]{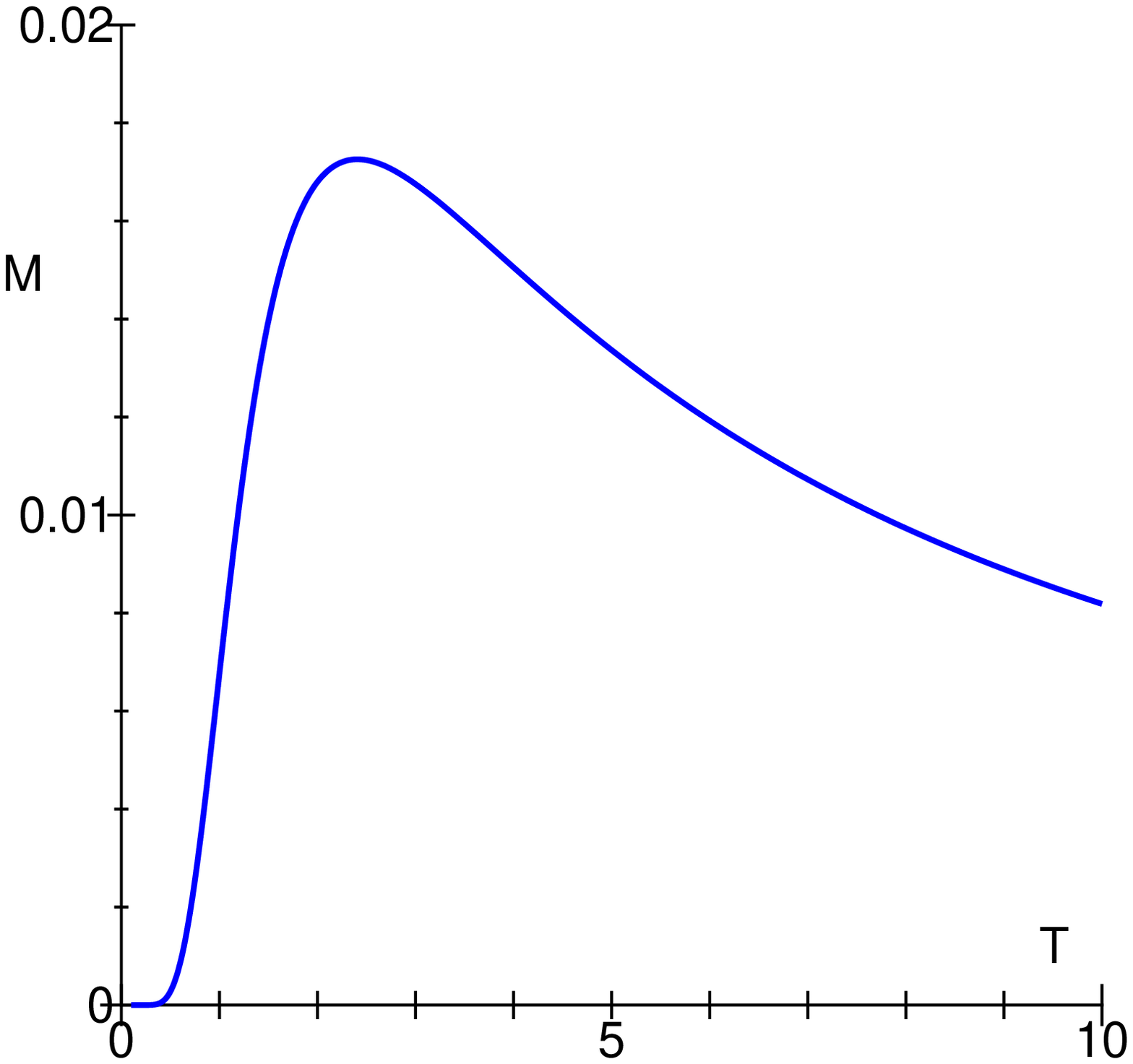}&
\includegraphics[height=35mm]{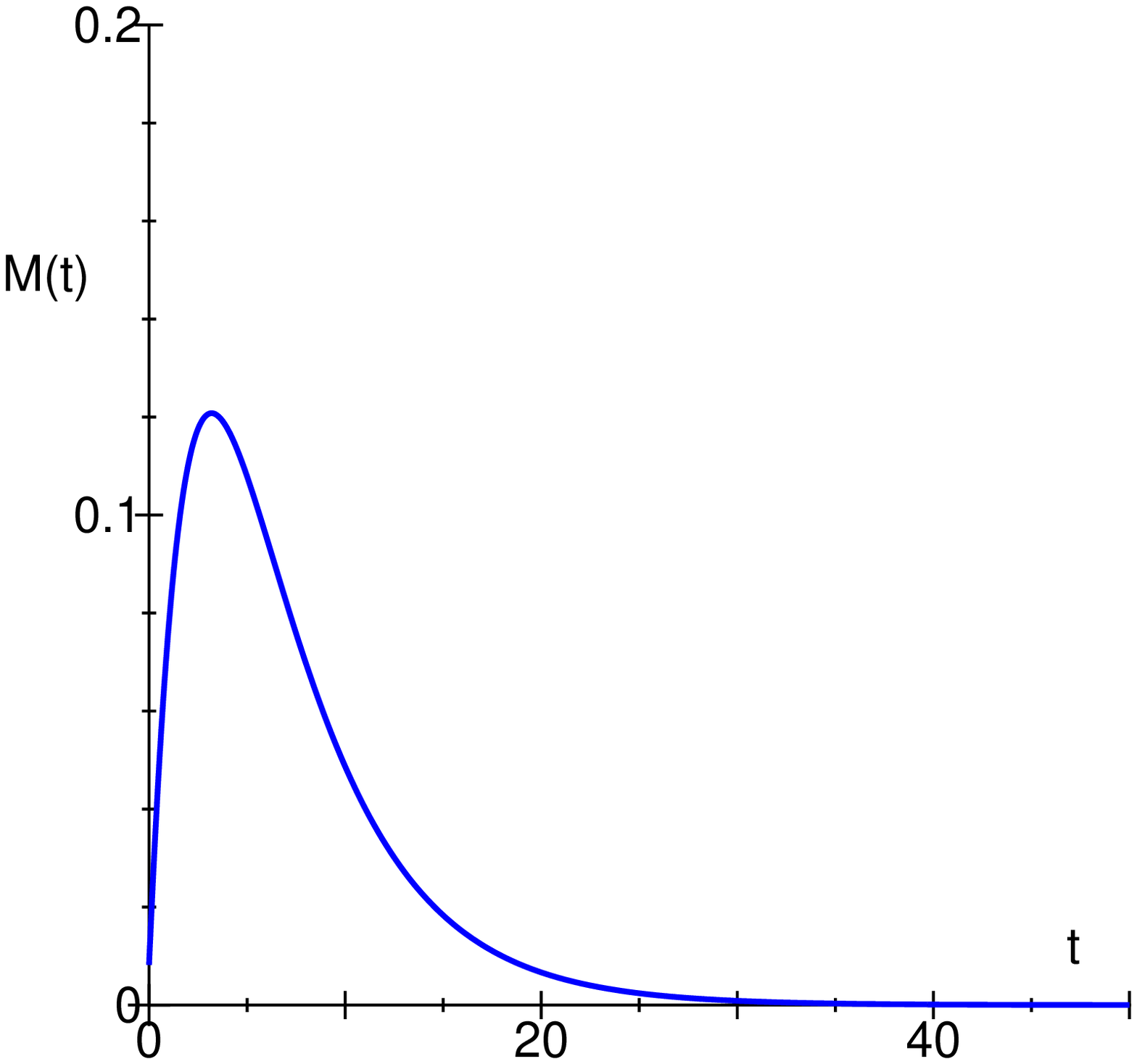}\\
({\rm a}) & ({\rm b})\end{array}$$
\caption{(a) Temperature dependence of magnetization $M$ of the spin antiferromagnetically coupled 
spin cluster (eq.~\ref{AF4}). (b) Time-evolution of $M$ after sudden quenching of the
temperature from $T=10$ to 0.1.}
\label{fig-AFpair}
\end{figure}

Next, we study the antiferromagnetically coupled spin cluster in a uniform magnetic field
\beq
{\cal H}=J(\sigma_1\sigma_2+\sigma_2\sigma_3+\sigma_3\sigma_4+\sigma_4\sigma_1)
-H(\sigma_1+\sigma_2+\sigma_3+\sigma_4),
\label{AF4}
\eeq
where $J$ and $H$ are competing. 
In other words, this system has ``frustration'' 
between the magnetic coupling and the magnetic field.
In Fig.~\ref{fig-AFpair}(a), we depict the temperature dependence
of the magnetization $M=\sigma_1+\sigma_2+\sigma_3+\sigma_4$ in the magnetic field $H=0.1J$. 
The magnetization shows nonmonotonic behavior as a function of temperature.
The temperature is scaled by $J$. The time evolution of $M$ after sudden quenching of the
temperature from $T=10$ to 0.1 is plotted in Fig.~\ref{fig-AFpair}(b).
Here we again find the same nonmonotonic behavior.

\begin{figure}[h]
$$
\includegraphics[height=35mm]{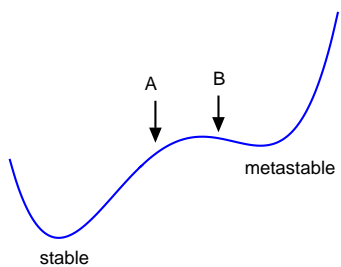}
$$
\caption{Schematic free energy structure with a metastable state.}
\label{metastable}
\end{figure}

We also refer to nonmonotonic behavior in macroscopic models.
If the system has a metastable state that is separated from the equilibrium state
by a free energy barrier, as depicted in Fig.~\ref{metastable}, the system can show 
nonmonotonic behavior. For example, if the initial state is point B,
it relaxes to the metastable point in a short time, and later it relaxes to the
equilibrium state via a nucleation process. This is characteristically different from
the simple relaxation starting from point A. Thus, this type of difference in the relaxation
has been used to detect a metastable state.\cite{konishi}
In the present macroscopic case, there are also two competing structures representing the metastable state
and the equilibrium state.

In the present study, we only investigated the cases with sudden changes in the temperature. 
It is also interesting to change the temperature with a finite sweep velocity.
If we sweep the temperature sufficiently slowly, a quasi-static state should be realized.
The sweep velocity dependence of the dynamics will be studied in the near future.

Using a lattice constructed by the decoration bond, we can study the
reentrant phase transition.\cite{vincent,tanaka}
If we change the temperature suddenly from the paramagnetic region to 
the ferromagnetic region, we obtain very slow relaxation due to an 
entropy-induced screening effect.\cite{tanaka}
Here, we find that the antiferromagnetic correlation appears for only a very short time and 
the nonmonotonicity does not develop.
To achieve nonmonotonic behavior on a macroscopic scale, we have to construct a system 
where the present mechanics takes place on a coarse-grained scale.
Here, the present spin $\sigma_i$ should represent a coarse-grained local magnetization. 
In spin glasses, the frustrated configuration remains in each step of the coarse-grain magnetization process,
which may be called a ``hierarchical structure''. 
Furthermore, in real spin glasses, the spin directions in the ordered configuration are spatially 
random, and we need to average these over all the possible configurations. Thus, the temperature
dependence of the uniform magnetization is averaged out, although the local spin correlations
may show various peculiar temperature dependences. 
However, the hierarchical frustrated structure
causes peculiar phenomena in the time dependence. As was proposed by Takayama and Hukushima, 
the magnetization first decreases while the short-range spin glass correlation still increases,
whereas it finally increases after the correlation reaches the saturated value where
the local magnetization behaves independently.  
This complexity causes the peculiar spin glass properties,
but the present mechanism may give the basic local mechanism of these phenomena.

\acknowledgement

We would like to express our thanks to Professor Takayama for the kind explanation of
the paper.\cite{takayama}
This work was partially supported by Grant-in-Aid for Scientific
Research on Priority Areas
``Physics of new quantum phases in superclean materials'' (Grant No.
17071011) from MEXT,
by the Next Generation Super Computer Project, Nanoscience
Program from MEXT,
and also by the 21st Century COE Program at the University of Tokyo 
``Quantum Extreme Systems and Their Symmetries''
from the Ministry of Education, 
Culture, Sports, Science and Technology of Japan.
The authors thank the Supercomputer Center, Institute for Solid State
Physics, University of Tokyo for the use of the facilities.


\begin{thebibliography}{99} 

\bibitem{takayama} 
H.~Takayama and K.~Hukushima: 
J. Phys. Soc. Jpn. {\bf 76} (2007) 013702.

\bibitem{Tjonsson} 
T.~Jonsson, K.~Jonason, and P.~Nordblad:
Phys. Rev. B {\bf 59} (1999) 9402.

\bibitem{PEjonsson}
P.~E.~J\"{o}nsson and H.~Takayama:
J. Phys. Soc. Jpn {\bf 74} (2005) 1131.

\bibitem{syozi} 
I.~Syozi: 
Phase Transition and Critical Phenomena, Vol. 1, 
Domb and Green eds. (Academic Press, New York, 1972) p.269.

\bibitem{miya1} 
S.~Miyashita: 
Prog. Theor. Phys. {\bf 69} (1983) 714.

\bibitem{kitatani} 
H.~Kitatani, S.~Miyashita, and M.~Suzuki: 
J. Phys. Soc. Jpn. {\bf 55} (1986) 865.

\bibitem{Jascur} 
M.~Ja\v{s}\v{c}ur and J.~Stre\v{c}ka: 
Condens. Matter Phys. {\bf 8} (2005) 869.

\bibitem{vincent} 
S.~Miyashita and E.~Vincent: 
Eur. Phys. J. B {\bf 22} (2001)  203.

\bibitem{tanaka} 
S.~Tanaka and S.~Miyashita: 
Prog. Theor. Phys. Suppl. {\bf 157} (2005) 34.

\bibitem{glauber}
R.~J.~Glauber: 
J. Math. Phys., {\bf 4} (1963) 294.

\bibitem{eigen} The time evolution function ${\cal L}$ is symmetrized in the form
${\hat P}_{\rm eq}^{-1/2}{\cal L}{\hat P}_{\rm eq}^{1/2}\equiv{\hat {\cal L}}$. 
Here, ${\hat P}_{\rm eq}^{1/2}$ is the diagonal matrix with 
$({\hat P}_{\rm eq}^{1/2})_{ii}=\sqrt{ P_{\rm eq}(i) }$.
Let $\{\mbold{v}_k \}$ be the eigenvectors of ${\hat {\cal L}}$, 
i.e., ${\hat {\cal L}}\mbold{v}_k=\lambda_k\mbold{v}_k$.  
The eigenvectors $\{\mbold{u}_k \}$ of ${\cal L}$ are given by ${\hat P}_{\rm eq}^{1/2}\mbold{v}$
with the same eigenvalues. 
Here we must be careful about the normalization of the vector.
As the eigenstate of the matrix $\hat{\mathcal{L}}$, 
the eigenstates are normalized as $\| \mathbf{v}_k \|=1$.
On the other hand, the normalization of the probability is given by $\sum_i P_{\mathrm{eq}} (i) = 1$
(*). Although $P_{\mathrm{eq}}^{1/2} v_1 = u_1 \propto \mathbf{P}_{\mathrm{eq}}$, 
we need to scale $\{ \mathbf{u}_k \}$ to satisfy the condition (*).

\bibitem{konishi} Y.~Konishi, H.~Tokoro, M.~Nishino, and S.~Miyashita: 
        J. Phys. Soc. Jpn. {\bf 75} (2006) 114603.

\end{thebibliography}
\end{document}